\documentclass[%
 reprint,
 amsmath,amssymb,
 aps,
]{revtex4-2}

\usepackage{graphicx,color}  
\usepackage{amsmath,amssymb}  
\usepackage{hyperref}  
\usepackage{physics}   
\usepackage{dcolumn}   
\usepackage[normalem]{ulem}
\usepackage{xcolor}

\begin{document}

\title{Entanglement between pair-created twin universes with opposite time arrows\\ should leave a birthmark on CMB spectrum}

\author{
Pisin Chen$^{a,b,c,d}$
}
\email{pisinchen@phys.ntu.edu.tw}

\author{
Kuan-Nan Lin$^{e}$
}
\email{knlinphy@gmail.com}

\author{
Wei-Chen Lin$^{f,g,h}$
}
\email{archennlin@gmail.com}

\author{
Dong-han Yeom$^{a,i,j,k}$
}
\email{innocent.yeom@gmail.com}


\affiliation{
$^{a}$Leung Center for Cosmology and Particle Astrophysics, National Taiwan University, Taipei 10617, Taiwan\\
$^{b}$Department of Physics and Center for Theoretical Sciences, National Taiwan University, Taipei 10617, Taiwan\\
$^{c}$Graduate Institute of Astrophysics, National Taiwan University, Taipei 10617, Taiwan\\
$^{d}$Kavli Institute for Particle Astrophysics and Cosmology,
SLAC National Accelerator Laboratory, Stanford University, Stanford, California 94305, USA\\
$^{e}$Asia Pacific Center for Theoretical Physics (APCTP), Pohang, 37673, Republic of Korea\\
$^{f}$Department of Physics, Pusan National University, Busan 46241, Republic of Korea\\
$^{g}$Extreme Physics Institute, Pusan National University, Busan 46241, Republic of Korea\\
$^{h}$Department of Science Education, Ewha Womans University, Seoul 03760, Republic of Korea\\
$^{i}$Department of Physics Education, Pusan National University, Busan 46241, Republic of Korea\\
$^{j}$Research Center for Dielectric and Advanced Matter Physics, Pusan National University, Busan 46241, Republic of Korea\\
$^{k}$Department of Physics and Astronomy, University of Waterloo, Waterloo, ON N2L 3G1, Canada
}


\begin{abstract}

Why (and how) the Universe was born is one of the ultimate questions in physics. Another big puzzle is about the arrow of time: why is there only one direction of time? Are these two issues related? One way to solve both puzzles at one stroke is to posit that our universe was pair-created with a twin, whose time arrow is opposite to ours. If so, then the twins must naturally be quantum entangled. In Euclidean quantum gravity, this implies the existence of a Euclidean wormhole bridging the twin universes. Each universe is then in a mixed-state and the mutual entanglement shall leave signatures in the cosmic microwave background (CMB) power spectrum. Invoking the Klebanov-Susskind-Banks wormhole as a toy model for the sake of tractability, we show that the entanglement selects a novel and unique global vacuum for the total inflaton perturbations in both universes. This is equivalent to imposing a simple harmonic oscillator boundary condition on the Euclidean wavefunction of the total perturbations, and it turns out that the entanglement enhances the CMB power spectrum for long-wavelength modes. Such a birthmark renders our notion refutable.

\end{abstract}

\maketitle


\textit{Introduction}---Why and how was the universe born? The standard Big Bang theory suggested the birth of the universe from a hot, dense spacetime singularity \cite{Hawking:1970zqf}. But why and how did that happen? There have been a plethora of solutions proposed, where physical entities beyond the big bang singularity, either by classical or quantum means, are invoked. To name just a few, the ekpyrotic model suggested to replace the hot Big Bang by a big bounce due to periodic brane collisions in higher dimensional spacetime \cite{khoury2001ekpyrotic} (See also the counter argument by Linde \cite{linde2002inflationary}). Loop quantum cosmology also supports the big bounce scenario via quantum geometry effects~\cite{Ashtekar:2008zu}. Conformal cyclic cosmology suggests that the big bang of each aeon (cosmic cycle) can be viewed as the future boundary of the previous aeon \cite{penrose2013cycles}. What is the evidence for any of these proposals?

Another big puzzle about the universe is the \textit{arrow of time}. The thermodynamic arrow of time can be addressed by thermodynamic entropy and the second law of thermodynamics \cite{page1984,carroll2004spontaneous}. However, if time had already emerged before the hot big bang, then the argument for the arrow of time based on thermodynamics becomes unclear. On the other hand, entanglement entropy seems to provide a fundamental quantum origin for the arrow of time \cite{chen2014evolution,bellido2021effects,al2024decoherent}. But what did the baby universe entangle with in the first place?

In this Letter, we attempt to address these puzzles via the scenario of pair-created twin universes with opposite time arrows. Euclidean quantum gravity allows twin universes to be created from nothing due to quantum fluctuations, and the twins are endowed with opposite signs of the time arrow, where the global time symmetry is preserved. We point out that quantum entanglement between the twin universes at birth would leave a birthmark on the cosmic microwave background (CMB) power spectrum.

It is well known that Euclidean gravity can facilitate the initial quantum state of a Lorentzian system, through which the cosmic microwave background (CMB) power spectrum is affected \cite{Halliwell:1984eu,Chen:2016ask,Chen:2017qeh,Chen:2019cmw,Chen:2024ckx}. In Hartle-Hawking's \textit{no-boundary proposal} (NBP) \cite{Hartle:1983ai}, a de Sitter universe is nucleated from a Euclidean hemisphere, which imposes strict boundary conditions at the pole such that the \textit{Bunch-Davies vacuum} is uniquely selected for the inflaton perturbations \cite{Laflamme:1987mx,Chen:2017aes,Chen:2019mbu}.

While not yet at the $5\sigma$-level precision, the current CMB observations hint on deviations of the long wavelength modes in the CMB spectrum from the prediction based on the conventional Bunch-Davies vacuum. On the theoretical side, it has been suggested that the boundary conditions imposed by the Hartle-Hawking proposal are unnecessarily restrictive.
Indeed, if the pole of the Hartle-Hawking hemisphere is replaced by a Euclidean wormhole, the boundary conditions can be relaxed and alternative quantum states deviating from that based on the Bunch-Davies vacuum naturally emerge \cite{Chen:2024ckx}.

When the twin universes are connected via a wormhole, the inflaton fluctuations in each universe must be entangled with the other. In order to have a well-defined notion of entanglement between them, we found that the quantum state of the total perturbations is uniquely determined by a novel \textit{global} vacuum. In terms of the wavefunction, it is equivalent to imposing the boundary condition that the Euclidean on-shell wavefunction of the total perturbation should \textit{behave like a simple harmonic oscillator} (SHO).

When focusing on one of the twin universes, the perturbation is a \textit{mixed state} with Bunch-Davies vacuum as its ground state. Based on that, we compute the resulting CMB power spectrum in one of the universes as well as the entanglement entropy between the two universes. We conclude that the long-wavelength modes in the spectrum would be enhanced and an entanglement entropy would emerge if our universe was indeed born with its twin and the two are bridged with a Euclidean wormhole, where the amount of enhancement depends on the wormhole size.

\textit{A simple harmonic oscillator toy model}---
Let us first review how to extract information of entanglement from the Euclidean action of a simple harmonic oscillator in a Euclidean flat space, which sets up the stage for our later study of cosmology. A SHO has the action:
\begin{eqnarray}\label{S-sho}
    S_{\mathrm{E}}
    =
    \frac{m}{2}\int_{\tau_{\mathrm{II}}}^{\tau_{\mathrm{I}}} d\tau \left[ \dot{x}^{2}(\tau)+\omega^{2}x^{2}(\tau) \right],
\end{eqnarray}
where $m$ is the mass, $x$ the position, $\omega$ the frequency, $\tau$ the Euclidean time, $\tau_{\mathrm{I},\mathrm{II}}$ the boundary points, and the dot denotes taking derivative with respect to the argument.

An on-shell solution is one that satisfies the equation of motion, $\ddot{x}-\omega^{2}x=0$, with boundary conditions $x(\tau_{\mathrm{I},\mathrm{II}})=x_{\mathrm{I},\mathrm{II}}$. We will consider $\tau_{\mathrm{I},\mathrm{II}}=\pm\beta/4$ to achieve a thermofield double state, in which $\beta$ can be interpreted as temperature \cite{laflamme1989geometry}. The solution is then expanded by
\begin{eqnarray}\label{x-sho}
    x(\tau)
    =
    A g(\tau)
    +
    B d(\tau),
\end{eqnarray}
where the basis functions are
\begin{eqnarray}\label{g,d-sho}
    g(\tau)
    =
    e^{\omega\tau}
    \quad
    \mathrm{and}
    \quad
    d(\tau)
    =
    e^{-\omega\tau},
\end{eqnarray}
and the coefficients $\{A,B\}$ are determined by the boundary conditions. The on-shell Euclidean action is then
\begin{eqnarray}\label{SE-sho}
    \frac{S_{\mathrm{E}}}{m\omega}
    =
    \frac{(1+z^2)(x_{\mathrm{I}}^2+x_{\mathrm{II}}^2)-4zx_{\mathrm{I}}x_{\mathrm{II}}}{2(1-z^2)},
    \;
    z=e^{-\beta\omega/2}.
\end{eqnarray}
By applying the Mehler formula, one obtains
\begin{equation}\label{wavefunction-sho}
\begin{aligned}
        e^{-S_{\mathrm{E}}}
        \propto& \sum_{\mu=0}^{\infty}e^{-\mu\beta\omega/2}\frac{e^{-\beta\omega/4}}{2^{\mu}\mu!}e^{-m\omega x_{\mathrm{I}}^2/2}
        \\
        & \times H_{\mu}\left(\sqrt{m\omega}x_{\mathrm{I}}\right)e^{-m\omega x_{\mathrm{II}}^2/2}H_{\mu}\left(\sqrt{m\omega}x_{\mathrm{II}}\right),
\end{aligned}
\end{equation}
where $H_{\mu}$ is the Hermite polynomial \cite{laflamme1989geometry}.

In canonical quantization where a system is partitioned into two subsystems, $\mathrm{I}$ and $\mathrm{II}$, the global vacuum $|\Omega\rangle$, which spans over both subsystems, is given by
\begin{eqnarray}
    |\Omega\rangle
    =
    \prod_{i}|\Omega_{i}\rangle
    =
    \prod_{i}\frac{1}{|\alpha_{i}|}\sum_{\mu=0}^{\infty}
    \left( \frac{\bar{\beta}_{i}}{\bar{\alpha}_{i}} \right)^{\mu}
    |\mu_{i}\rangle_{\mathrm{I}}|\mu_{i}\rangle_{\mathrm{II}},
\end{eqnarray}
where $\{\alpha_{i},\beta_{i}\}$ are Bogoliubov coefficients in mode $i$, bar denotes taking complex conjugate, $\mu=0$ corresponds to the local vacuum state restricted to each subsystem, and $\mu=1,2,\cdots$ are the local excited states. This shows that $|\Omega\rangle$ is an entangled two-mode pure state and the entanglement entropy between the two subsystems is
\begin{eqnarray}\label{entropy-sho}
    S_{\mathrm{EE}}
    =
        \sum_{i}
        \left(
        \ln|\alpha_{i}|^{2}
        -
        |\beta_{i}|^{2}\ln\left|\frac{\beta_{i}}{\alpha_{i}}\right|^{2}
        \right).
\end{eqnarray}

Acting the boundary conditions $\langle x_{\mathrm{II}}|\langle x_{\mathrm{I}}|$ on $|\Omega\rangle$ and comparing the result with Eq.~\eqref{wavefunction-sho}, one identifies $e^{-\beta\omega/2}=|\beta_{\omega}/\alpha_{\omega}|$ and $H_{\mu}$'s as the wave functions in each subsystem, while $e^{-S_{\mathrm{E}}}$ is the wave function of the total system.

\textit{Euclidean instantons vs. global modes}---
It is even possible to construct Lorentzian mode functions from the Euclidean bases in Eq.~\eqref{g,d-sho} to recover the mode expansion of a position operator in canonical quantization. Suppose the Euclidean flat space in Eq.~\eqref{S-sho} is connected to the $\mathrm{I}$ and $\mathrm{II}$ Lorentzian spacetimes at the junctions $\tau_{\mathrm{I},\mathrm{II}}$ by the analytic continuations: $\tau\rightarrow -it_{\mathrm{I},\mathrm{II}}+\tau_{\mathrm{I},\mathrm{II}}$, then the two global mode functions in Lorentzian spacetimes are
\begin{eqnarray}
        U^{(1)}
        &=&
        N_{1}
        \left[ g( -it_{\mathrm{I}}+\tau_{\mathrm{I}})
        +
        g( -it_{\mathrm{II}}+\tau_{\mathrm{II}})\right]
        \\
        &=&
        N_{1}e^{\beta\omega/4}\left[e^{-i\omega t_{\mathrm{I}}}+e^{-\beta\omega/2}e^{-i\omega t_{\mathrm{II}}}\right],\label{U-sho-1}
        \\
        U^{(2)}
        &=&
        N_{2}
        \left[d(-it_{\mathrm{II}}+\tau_{\mathrm{II}})
        +
        d(-it_{\mathrm{I}}+\tau_{\mathrm{I}})\right]
        \\
        &=&
        N_{2}e^{\beta\omega/4}\left[e^{i\omega t_{\mathrm{II}}}+e^{-\beta\omega/2}e^{i\omega t_{\mathrm{I}}}\right],\label{U-sho-2}
\end{eqnarray}
where $N_{1,2}=\mathrm{const.}$ In fact, Eqs.~\eqref{U-sho-1} and \eqref{U-sho-2} have identical structure as the global modes constructed by Unruh \cite{unruh1976notes}, where $e^{-\beta\omega/2}=\beta_{\omega}/\alpha_{\omega}$ and the global modes are superposition of local positive and negative frequency modes, but now with the \textit{feature of their connection with the Euclidean bases} \eqref{g,d-sho}. Finally, the global vacuum is defined by the global modes $\{U^{(1)},U^{(2)}\}$, while the local vacua are defined by the local modes $\{e^{-i\omega t_{\mathrm{I}}},e^{i\omega t_{\mathrm{II}}}\}$.

\begin{figure}[h]
    \centering
    \includegraphics[width=\linewidth]{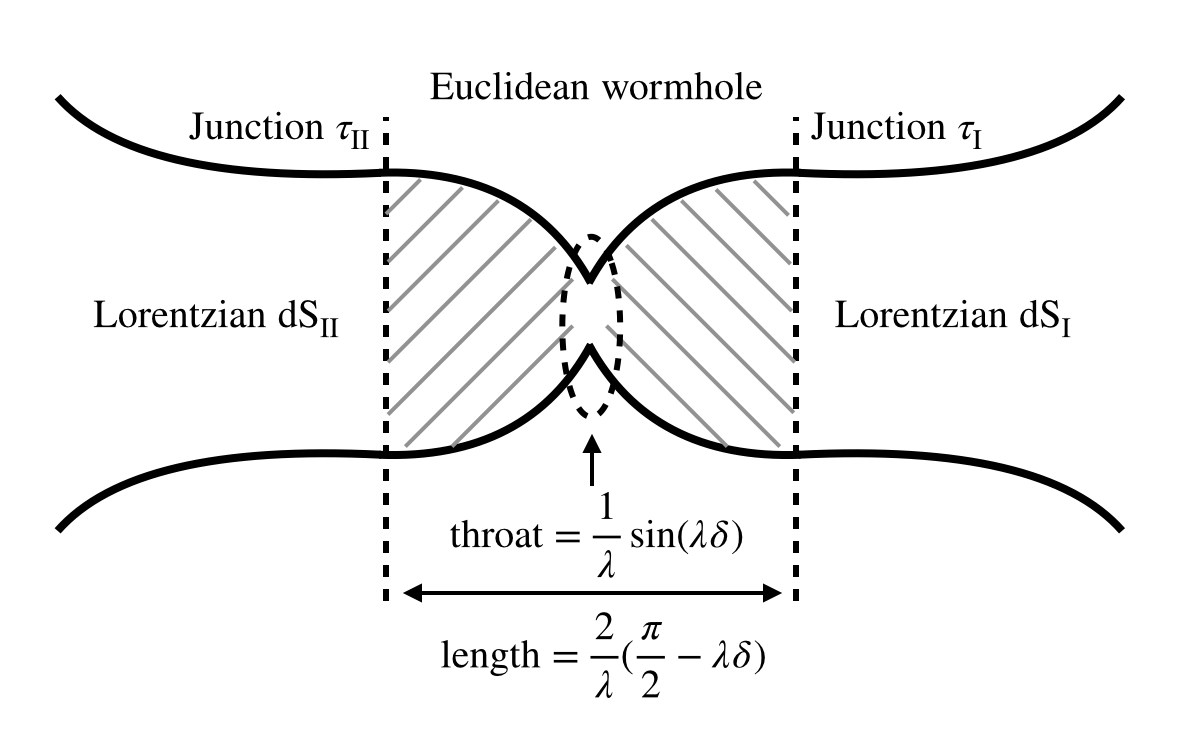}  
    \caption{Universes I and II are connected by a wormhole at $\dot{a}(\tau_{\mathrm{I,II}})=0$. Depending on the model, the throat may be smoothed out, however, at the cost of losing analytic analysis.} 
    \label{fig-wormhole}
\end{figure}

\textit{Euclidean wormhole}---
We now consider the Klebanov-Susskind-Banks (KSB) wormhole \cite{klebanov1989wormholes}:
\begin{eqnarray}
    ds^{2}
    =
    \sigma^{2} \left( d\tau^{2}
    +
    a^{2}(\tau)d\Omega_{3}^{2} \right),
\end{eqnarray}
where $d\Omega_{3}^{2}$ is the line element on a unit 3-sphere, and the Euclidean scale factor is $a(\tau)=\lambda^{-1}\sin \left[\lambda(|\tau|+\delta)\right]$, where $\{a,\tau,\lambda,\delta\}$ are dimensionless, $\sigma^{2}=2l_{\mathrm{P}}^{2}/(3\pi)$, $l_{\mathrm{P}}$ is the Planck length, and $\lambda=\sigma\sqrt{\Lambda/3}$, where $\Lambda$ is the standard cosmological constant for inflation \cite{Halliwell:1984eu}. (To convert back to real-world observations, one simply replaces $\sigma$ by the actual comoving curvature radius $R_{c}$ of the universe.) Evidently, $\delta$ is related to the existence of a wormhole (see Fig.~\ref{fig-wormhole}). We will consider the wormhole connecting closed dS universes $\mathrm{I}$ and $\mathrm{II}$ at the junctions $\tau_{\mathrm{I},\mathrm{II}}=\pm\pi/(2\lambda)\mp\delta$ and perform similar analytic continuations as before.

Let us consider the inflaton perturbation $f(\tau)$ in this scenario. Its on-shell Euclidean action is given by \cite{Laflamme:1987mx}
\begin{eqnarray}\label{S-ew}
    S_{\mathrm{E}}
    =
    \left.\frac{1}{2}a^3\dot{f}f\right|_{\tau_{\mathrm{II}}}^{\tau_{\mathrm{I}}},
\end{eqnarray}
and it satisfies the equation of motion:
\begin{eqnarray}\label{f-eom}
    \ddot{f}
    +
    3\frac{\dot{a}}{a}\dot{f}
    -
    \left(\frac{n^2-1}{a^2} \right)f
    =
    0,
\end{eqnarray}
for a massless perturbation (for simplicity), and $n$ is the mode number introduced by the scalar harmonics $Q_{n\ell m}$.

In $\tau\geq 0$, there are two linearly independent solutions:
\begin{eqnarray}\label{g,d-wh}
    g(\tau)&=&
    C\frac{\left[n+\cos(\lambda(\tau+\delta))\right]\sin^{n-1}(\lambda(\tau+\delta))}{\cos^{2n}(\lambda(\tau+\delta)/2)},
    \\ \label{g,d-wh2}
    d(\tau)&=&
    \bar{C}\frac{\left[n-\cos(\lambda(\tau+\delta))\right]\sin^{n-1}(\lambda(\tau+\delta))}{\sin^{2n}(\lambda(\tau+\delta)/2)},
\end{eqnarray}
where $C=\mathrm{const.}$ With Eqs.~\eqref{g,d-wh} and \eqref{g,d-wh2}, the perturbation in $\tau\in[\tau_{\mathrm{II}},\tau_{\mathrm{I}}]$ permits the general solution:
\begin{eqnarray}\label{f-ew}
    f(\tau)
    =
    AG(\tau)
    +
    BD(\tau),
\end{eqnarray}
where the basis functions are
\begin{eqnarray}
        G(\tau)
        &=&
        \Theta(\tau)rg(\tau)
        +
        \Theta(-\tau)r^{-1}d(-\tau),
        \\
        D(\tau)
        &=&
        \Theta(\tau)r^{-1}d(\tau)
        +
        \Theta(-\tau)rg(-\tau),
\end{eqnarray}
where $r$ is a constant. These bases map the current cosmology model to that of an SHO. By imposing the boundary conditions: $f(\tau_{\mathrm{I}})=f_{\mathrm{I}}$ and $f(\tau_{\mathrm{II}})=f_{\mathrm{II}}$, one obtains
\begin{eqnarray}\label{SE-ew}
        S_{\mathrm{E}}
        =
        \frac{(1+z^2)(F_{\mathrm{I}}^2+F_{\mathrm{II}}^2)-4zF_{\mathrm{I}}F_{\mathrm{II}}}{2(1-z^2)},
\end{eqnarray}
which has the same form as Eq.~\eqref{SE-sho}, where
\begin{align*}
        z=\frac{1}{r_{n}^2}
        =\frac{1}{r^{2}e^{in\pi}},
        \quad
        F_{\mathrm{I},\mathrm{II}}=\sqrt{\frac{\lambda a^{3}(\tau_{\mathrm{I,II}})(n^2-1)}{n}}f_{\mathrm{I},\mathrm{II}}.
\end{align*}
For a convergent wave function, $z^{2}\leq 1$. It follows from the SHO discussion that the global mode functions constructed from $\{G,D\}$ define a global vacuum that is related to the Bunch-Davies vacuum with the coefficients: $\beta_{n}/\alpha_{n}=1/r_{n}^{2}$, where $r_{n}$ can be determined from the continuity conditions $G(0_{+})=G(0_{-})$ and $D(0_{+})=D(0_{-})$:
\begin{eqnarray}\label{rn}
    \frac{1}{r_{n}^2}
    =
    \left(\frac{n+\cos \lambda\delta}{n-\cos\lambda\delta}\right)\tan^{2n} \left(\frac{\lambda\delta}{2}\right).
\end{eqnarray}
By using $|\alpha_{n}|^{2}-|\beta_{n}|^{2}=1$, one obtains $|\beta_{n}|^{2}=1/(r_{n}^{4}-1)$. It then follows that $z^{2}\leq 1$ is equivalent to $|\beta_{n}|^{2}\geq 0$.

\textit{Inflaton power spectrum}---
The observed inflaton power spectrum in our universe is a precious window to reveal the nature of the Big Bang. It is given by
\begin{eqnarray}\label{Pn-ew}
        P(n)
        =
        P_{\mathrm{BD}}(n)\left(1+2|\beta_{n}|^{2}\right),
\end{eqnarray}
which is obtained from the reduced density matrix of the perturbation restricted to, say, Universe $\mathrm{I}$. The non-vanishing $|\beta_{n}|^{2}$ leads to an enhancement in the spectrum for small $n$, while the spectrum based on the standard Bunch-Davies vacuum in closed dS gives the suppression:
\begin{eqnarray}
    P_{\mathrm{BD}}(n)
    =
    \frac{\lambda^{2}}{4\pi^{2}}\frac{n^{2}(1+y^{2})}{n^{2}+y^{2}},
\end{eqnarray}
where $y=n/\sinh(\lambda t_{\mathrm{I}})$. In our scenario, $P(n)$ at small $n$ depends on the competition between the pair-universe entanglement and the dS curvature. For sufficiently large $n$, the standard notion of scale invariance is recovered. Interestingly, in $1\ll n\ll 1/q$, where $q=\pi/2-\lambda\delta$, an effective thermal interpretation emerges, i.e., $|\beta_{n}/\alpha_{n}|\simeq e^{-2nq}\simeq 1-2nq$, and
\begin{eqnarray}\label{P-thermal}
    P(1\ll n\ll 1/q)
    \simeq
    \frac{1}{2nq}\frac{\lambda^{2}}{4\pi^{2}}\frac{n^{2}(1+y^{2})}{n^{2}+y^{2}}.
\end{eqnarray}

\begin{figure}[h]
    \centering
    \includegraphics[width=\linewidth]{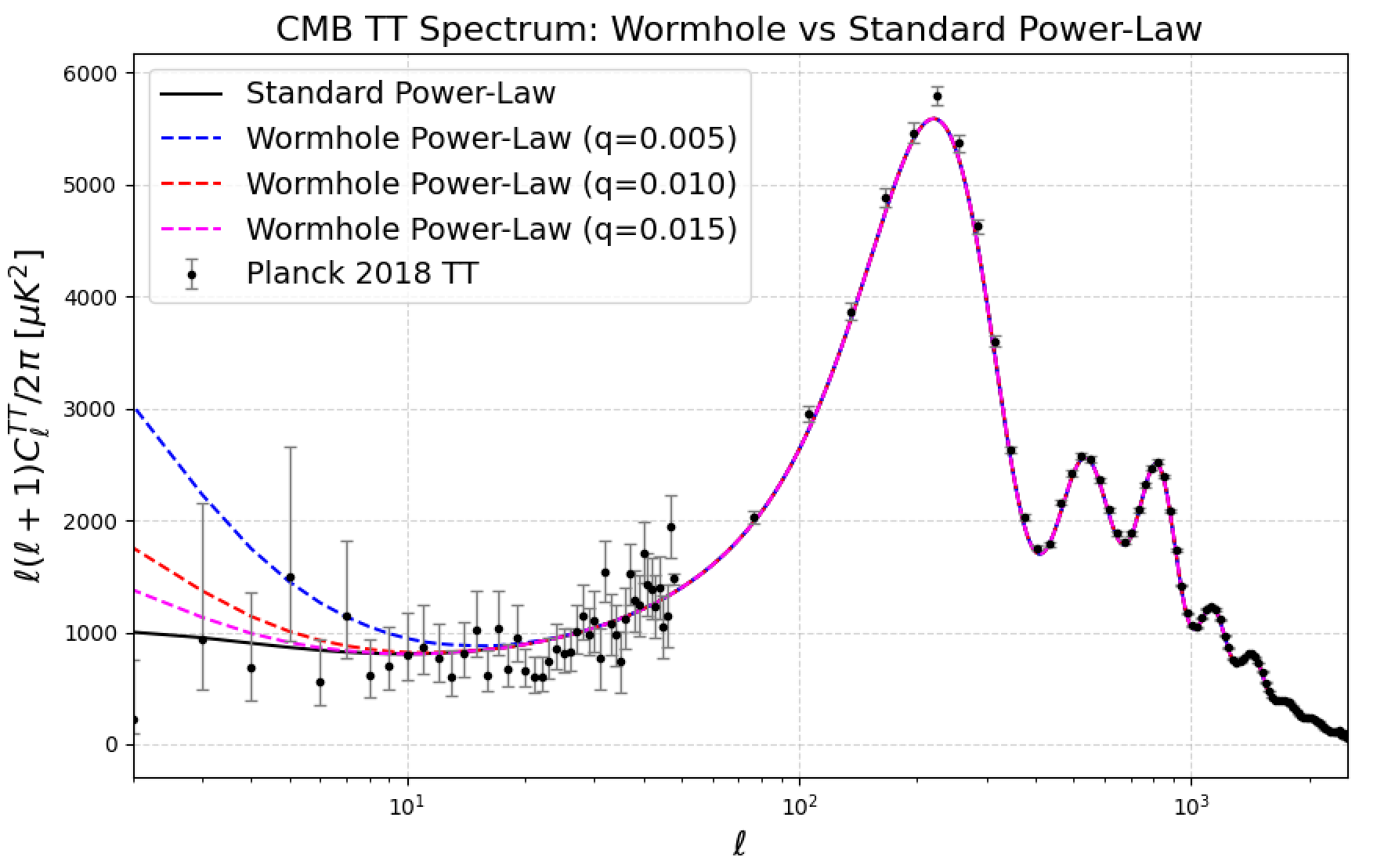}  
    \caption{Modification of normalized temperature anisotropy spectrum due to the entanglement between the twin universes based on Eq.~\eqref{Pn-ew}. Here, we assume that the universe was initially nucleated as dS, and after the horizon crossing ($y\ll 1$), it transitioned to power-law expansion. Since the entanglement originated \textit{before} inflation, $\lambda$ in Eq.~\eqref{rn} should be determined by the cosmological constant before the transition, while $P_{\mathrm{BD}}$ acquires time-dependence due to the power-law expansion after the transition. (For $n\gg 1$, $l\sim kd_{\mathrm{LSS}}\sim (d_{\mathrm{LSS}}/R_{c})n\sim 0.1 n$.) We used CAMB \cite{lewis2011camb} for the calculation.}
    \label{fig-spectrum-all}
\end{figure}

The competition between enhancement and suppression also appears in, e.g., massive inflaton in NBP \cite{Chen:2017qeh}, modified Starobinsky model \cite{Chen:2019cmw}, etc. Currently, the longest observable perturbation in the CMB anisotropies corresponds to the comoving wavenumber $k\sim 2\pi/2d_{\mathrm{LSS}}\simeq 0.23\;\mathrm{Gpc}^{-1}$, where $d_{\mathrm{LSS}}\simeq 13.8\;\mathrm{Gpc}$ is the current comoving angular diameter distance to the last scattering surface (LSS). This corresponds to the multipole moment $l\sim kd_{\mathrm{LSS}}\simeq 3$. Based on the current Hubble parameter $H_{0}=67.4\;\mathrm{km/(s\cdot Mpc)}$ and the curvature density parameter saturating the observational bound 
$\Omega_{K,0}\simeq -0.0012$ \cite{Planck2018}, $R_{c}=1/(\sqrt{|\Omega_{K,0}|}H_{0})\simeq 128.5\;\mathrm{Gpc}$. From $kR_{c}=\sqrt{n^2-1}$, the longest observable perturbation in CMB today then corresponds to $n\sim 29$. Typically, this perturbation only indicates that inflation should last at least $\sim 60$ e-folds, so it is possible that this perturbation did not exit the horizon during the early inflation but at a later time. One may then question whether quantum gravity effects persist if the perturbation did not exit the horizon early enough? In fact, the effect can persist and be observable today if the wormhole $(\lambda\delta)$ is not too small. For example, for $q=0.015$, the spectrum at $n\sim 29$, i.e., $l\sim 3$ today, is enhanced by a factor of $1.15$ based on Eq.~\eqref{P-thermal} (exact value is $1.43$ based on Eqs.~\eqref{rn} and \eqref{Pn-ew}) and the effect diminishes before reaching $n\sim 100$, i.e., $l\sim 11$, which are compatible with the observed CMB data $C_{l\geq 2}^{\mathrm{TT}}$ \cite{Planck2018} (see Fig.~\ref{fig-spectrum-all}). By taking the inflation scale as $\rho_{\mathrm{inf}}=\Lambda/8\pi l_{\mathrm{P}}^2\sim (10^{16}\;\mathrm{GeV})^{4}$, $q=0.015$ infers the physical throat radius $R_{c}/\lambda\sim 5\times 10^{5} l_{\mathrm{P}}$ and length $R_{c}(\tau_{\mathrm{I}}-\tau_{\mathrm{II}})\sim 1.5\times 10^{4} l_{\mathrm{P}}$ of a Euclidean wormhole.

Finally, for smaller wormholes, the enhancement is carried by even longer perturbations (smaller $n$'s) that may exit the horizon during the early inflation. Hence, their effects may not be observable today, but in the future.


\begin{figure}[h]
    \centering
    \includegraphics[width=\linewidth]{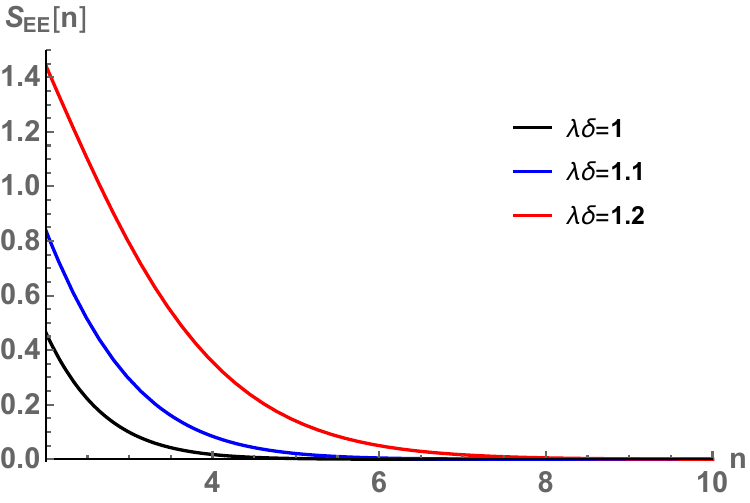}  
    \caption{Entanglement entropy between perturbations across a Euclidean wormhole. The larger $\lambda\delta$ is, the shorter the wormhole's length $(\tau_{\mathrm{I}}-\tau_{\mathrm{II}})$, the more the entanglement.  }
    \label{fig-entropy}
\end{figure}

\textit{Entanglement and the size of  wormhole}---
Although the entanglement entropy associated with the pair-universe may not be as relevant as the power spectrum observation-wise, the theoretical framework that we have developed here may play a significant role in the study of the black hole information paradox, which we leave for future work. Figure~\ref{fig-entropy} illustrates the entanglement entropy between pair-created inflaton perturbations in each universe according to Eq.~\eqref{entropy-sho}:
\begin{eqnarray}
        S_{\mathrm{EE}}
        =
        \sum_{n\ell m}^{\infty}S_{\mathrm{EE}}(n,\ell,m)
        =
        \sum_{n=2}^{\infty}S_{\mathrm{EE}}(n).
\end{eqnarray}

Figure~\ref{fig-entropy} shows that perturbations with larger $n$ are less entangled than those with smaller ones, since the reduced state in each universe is approximately the Bunch-Davies vacuum, i.e., $|\Omega_{n}\rangle\sim|0_{n};\mathrm{BD}\rangle_{\mathrm{I}}|0_{n};\mathrm{BD}\rangle_{\mathrm{II}}$ for large $n$. In addition, the wider and shorter the wormhole is, the stronger the entanglement. If the wormhole has a zero throat, i.e., ($\lambda\delta= 0$), then Universe $\mathrm{I}$ and Universe $\mathrm{II}$ reduce to two disconnected Hartle-Hawking no-boundary instantons, which are not entangled.

\textit{Conclusions}---
We have investigated the role of Euclidean wormholes that create paired universes. Inflaton perturbations in these Euclidean wormholes will propagate in both universes. If one traces the Bunch-Davies mode in one universe backward, then it will penetrate the Euclidean wormhole and propagate into the partner universe. Here, \textit{the Bunch-Davies mode of one universe \cite{footnote} corresponds to a non-Bunch-Davies mode in its partner universe}, and the relative amplitude $|\beta_{n}/\alpha_{n}|$ depends on the radius and the length of the Euclidean wormhole, which in turn depends on $\{\lambda,\delta\}$. By imposing the boundary condition that the wave function of the perturbation behaves as a SHO, Eq.~\eqref{SE-sho} vs. Eq.~\eqref{SE-ew}, we show that our universe begins as a mixed state, where the power spectrum and the entanglement entropy between the two universes are computed in the usual way.

The most striking result of this work is the implication of the wormhole in the power spectrum: the enhancement of long-wavelength modes due to pair-universe entanglement, while the standard scale-invariance for short-wavelength modes is preserved. The magnitude of the enhancement is determined by the feature of the wormhole and the spatial curvature, which affects the mapping between $n$ and $l$. By comparing our prediction with observational data, the information about the wormhole may be revealed. 

Although we employ a wormhole toy model for analytic demonstration, the characteristic features of our results should remain qualitatively valid for generic models of wormholes, which may require numerical treatments. In addition, since the origin of the entanglement occurred \textit{before} inflation, we argue that the enhancement factor $2|\beta_{n}|^{2}$ in the power spectrum should remain time-independent even if the nucleated de Sitter inflation subsequently transitions to slow-roll. That is, different subsequent slow-roll scenarios should only modify $P_{\mathrm{BD}}$, thus the wormhole entanglement enhancement remains intact. 

Let us address our notion more generically: \textit{the possible observation of non-Bunch-Davies modes would be an evidence that our universe is entangled with a partner universe.} Non-Bunch-Davies modes cannot be induced by compact instantons, whereas a wormhole can. There may be alternative explanations for the existence of non-Bunch-Davies modes. However, if we assume that our universe was born together with its twin universe and they were connected, which is inevitable from the Euclidean path integral point of view, then the pair universes form a ground state quite naturally, from which the non-Bunch-Davies modes are induced. These notions render our proposal refutable through future CMB observations.

\textit{Acknowledgements}---
PC was supported by Taiwan’s National Science and Technology Council (NSTC) under project number 110-2112-M-002-031, and by the Leung Center for Cosmology and Particle Astrophysics (LeCosPA), National Taiwan University (NTU). KL was supported by the Elite Doctoral Scholarship provided by NTU and NSTC, and by the YST Program at the Asia Pacific Center for Theoretical Physics (APCTP). DY was supported by the National Research Foundation of Korea (Grant No.: 2021R1C1C1008622, 2021R1A4A5031460). WL was supported by the National Research Foundation of Korea (Grant No.: 2021R1C1C1008622, 2021R1A4A5031460, and Basic Science Research Program; Grant No.: RS-2024-00336507).


\nocite{*}


\begin{thebibliography}{99}

\bibitem{Hawking:1970zqf}
S.~W.~Hawking and R.~Penrose,
Proc. Roy. Soc. Lond. A \textbf{314}, 529-548 (1970).

\bibitem{khoury2001ekpyrotic}
J.~Khoury, B.~A.~Ovrut, P.~J.~Steinhardt and N.~Turok,
Phys. Rev. D \textbf{64}, 123522 (2001).

\bibitem{linde2002inflationary}
A.~Linde,
\textit{Inflationary theory versus ekpyrotic/cyclic scenario},
arXiv preprint hep-th/0205259.

\bibitem{Ashtekar:2008zu}
A.~Ashtekar,
Gen. Rel. Grav. \textbf{41}, 707-741 (2009)
[arXiv:0812.0177 [gr-qc]].

\bibitem{penrose2013cycles}
R.~Penrose,
\textit{Cycles of time: an extraordinary new view of the universe}, Random House, 2013.













\bibitem{page1984}
D.~N.~Page,
Int. J. Theor. Phys. \textbf{23}, 725 (1984).

\bibitem{carroll2004spontaneous}
S.~M.~Carroll and J. Chen,
\textit{Spontaneous Inflation and the Origin of the Arrow of Time},
arXiv preprint hep-th/0410270.



\bibitem{chen2014evolution}
P.~Chen, P.-S.~Hsin and Y.~Niu,
JCAP \textbf{2014}(02), 040 (2014).


\bibitem{bellido2021effects}
S.~B.~Bellido,
Phys. Rev. D \textbf{104}, 106009 (2021).




\bibitem{al2024decoherent}
J.~Al-Khalili and E.~K.~Chen,
Found. Phys. \textbf{54}(4), 49 (2024).




\bibitem{Halliwell:1984eu}
J.~J.~Halliwell and S.~W.~Hawking,
Phys. Rev. D \textbf{31}, 1777 (1985).


\bibitem{Chen:2016ask}
P.~Chen, Y.~C.~Hu and D.~Yeom,
JCAP \textbf{07}, 001 (2017)
[arXiv:1611.08468 [gr-qc]].

\bibitem{Chen:2017qeh}
P.~Chen and D.~Yeom,
Eur. Phys. J. C \textbf{78}, no.10, 863 (2018)
[arXiv:1706.07784 [gr-qc]].

\bibitem{Chen:2019cmw}
P.~Chen, D.~Ro and D.~Yeom,
Phys. Dark Univ. \textbf{28}, 100492 (2020)
[arXiv:1904.00199 [gr-qc]].

\bibitem{Chen:2024ckx}
P.~Chen, K.~N.~Lin, W.~C.~Lin and D.~Yeom,
Phys. Rev. D \textbf{111}, 083520 (2025)
[arXiv:2404.15450 [gr-qc]].




\bibitem{Hartle:1983ai}
J.~B.~Hartle and S.~W.~Hawking,
Phys. Rev. D \textbf{28}, 2960-2975 (1983).





\bibitem{Laflamme:1987mx}
R.~Laflamme,
Phys. Lett. B \textbf{198}, 156-160 (1987).

\bibitem{Chen:2017aes}
P.~Chen, Y.~H.~Lin and D~Yeom,
Eur. Phys. J. C \textbf{78}, no.11, 930 (2018)
[arXiv:1707.01471 [gr-qc]].

\bibitem{Chen:2019mbu}
P.~Chen, H.~H.~Yeh and D.~Yeom,
Phys. Dark Univ. \textbf{27}, 100435 (2020)
[arXiv:1903.12045 [gr-qc]].




\bibitem{laflamme1989geometry}
R.~Laflamme,
Nucl. Phys. B \textbf{324}, 233-252 (1989).

\bibitem{unruh1976notes}
W.~G.~Unruh,
Phys. Rev. D \textbf{14}, 870 (1976).

\bibitem{klebanov1989wormholes}
I.~Klebanov, L.~Susskind and T.~Banks
Nucl. Phys. B \textbf{317}, 665-692 (1989).



\bibitem{Planck2018}
N. Aghanim et al.,
A $\&$ A \textbf{641}, A6 (2020).

\bibitem{lewis2011camb}
A.~Lewis, A.~Challinor and A.~Lasenby,
Astrophys. J. \textbf{538}, 473 (2000); 
Code for Anisotropies in the Microwave Background,
A.~Lewis and A.~Challinor, https://camb.info/

\bibitem{footnote}
Bunch-Davies mode refers to the analytic continuation of Eq.~\eqref{g,d-wh}, while the analytic continuation of Eq.~\eqref{g,d-wh2} is the complex conjugate of the Bunch-Davies mode.

\end{thebibliography}


\end{document}